%% file: main.tex
\documentclass[twocolumn]{article}

\usepackage{geometry}
\usepackage{graphicx}
\usepackage{amsmath} 
\usepackage{amsfonts}
\usepackage{amssymb} 
\usepackage{hyperref}
\usepackage[numbers,sort&compress]{natbib} 

\usepackage{authblk}
\usepackage{textcomp}

\title{Robust electron counting for direct electron detectors with the Back-Propagation Counting method}

\author[1]{Joshua Renner}
\author[2]{Matthew A. Wright}
\author[2]{Kristofer Bouchard\thanks{Corresponding authors. Emails: kebouchard@lbl.gov, percius@lbl.gov, agoldschmidt@lbl.gov}}
\author[3,4]{Bruce E. Cohen}
\author[3]{Peter Ercius\textsuperscript{\textasteriskcentered}}
\author[5]{Azriel Goldschmidt\textsuperscript{\textasteriskcentered}}
\author[3]{Cassio C.S. Pedroso}
\author[3]{Ambarneil Saha}
\author[3]{Peter Denes}

\affil[1]{Instituto de F\'{i}sica Corpuscular (IFIC), CSIC \& Universitat de Val\`{e}ncia, Calle Catedr\'{a}tico Jos\'{e} Beltr\'{a}n, 2, 46980 Paterna, Valencia, Spain}
\affil[2]{Scientific Data Division, Lawrence Berkeley National Laboratory, 1 Cyclotron Road, Berkeley, CA, USA, 94720}
\affil[3]{Molecular Foundry, Lawrence Berkeley National Laboratory, 1 Cyclotron Road, Berkeley, CA, USA, 94720}
\affil[4]{Division of Molecular Biophysics \& Integrated Bioimaging, Lawrence Berkeley National Laboratory, 1 Cyclotron Road, Berkeley, CA, USA, 94720}
\affil[5]{Engineering Division, Lawrence Berkeley National Laboratory, 1 Cyclotron Road, Berkeley, CA, USA, 94720}

\date{\today}

\begin{document}

\maketitle

\begin{abstract}
Electron microscopy (EM) is a foundational tool for directly assessing the structure of materials.
Recent advances in direct electron detectors have improved signal-to noise ratios via single-electron counting. However, accurately counting electrons at high fluence remains challenging. We developed a new method of electron counting for direct electron detectors, Back-Propagation Counting (BPC). BPC uses machine learning techniques designed for mathematical operations on large tensors but does not require large training datasets. In synthetic data, we show BPC is able to count multiple electron strikes per pixel and is robust to increasing occupancy. In experimental data, frames counted with BPC are shown to reconstruct diffraction peaks corresponding to individual nanoparticles with relatively higher intensity and produce images with improved contrast when compared to a standard counting method. Together, these results show that BPC excels in experiments where pixels see a high flux of electron irradiation such as \textit{in situ} TEM movies and diffraction.

\end{abstract}

\input{introduction}
\input{methods}
\input{results}
\input{discussion}

\vspace{1cm}
{\noindent\Large{\textbf{Code availability}}}\\\\
\noindent The code for the BPC method is available on GitHub at: \href{https://github.com/jerenner/backpropcount}{https://github.com/jerenner/backpropcount}.

\vspace{1cm}
{\noindent\Large{\textbf{Data availability}}}\\\\
\noindent The 4D-STEM datasets analyzed during the current study are not publicly available due to their large size but are available from the corresponding authors on reasonable request.

\bibliographystyle{naturemag}
\bibliography{bibliography}

\vspace{1cm}
{\noindent\Large{\textbf{Acknowledgments}}}\\\\
\noindent This work was primarily funded by the US Department of Energy in the program ``4D Camera Distillery: From Massive Electron Microscopy Scattering Data to Useful Information with AI/ML.'' JR acknowledges support from the Generalitat Valenciana of Spain under grant CIDEXG/2023/16. This research used resources of the National Energy Research Scientific Computing Center (NERSC), a Department of Energy Office of Science User Facility using NERSC award (BES-ERCAP0032504). Work at the Molecular Foundry was supported by the Office of Science, Office of Basic Energy Sciences, of the U.S. Department of Energy under Contract No. DE-AC02-05CH11231. This work was in part supported by the Laboratory Directed Research and Development Program of Lawrence Berkeley National Laboratory under U.S. Department of Energy Contract No. DE-AC02-05CH11231. 

\vspace{1cm}
{\noindent\Large{\textbf{Author contributions}}}\\\\
\noindent J.R. and A.G. devised the BPC counting method and J.R. implemented it in Python. J.R. wrote the initial draft of the manuscript and led revisions. A.G., J.R., K.B., M.W., and P.E. identified test cases for the method and contributed to the manuscript. M.W. provided PyTorch and ML expertise. P.E. devised the experiments carried out in the study and A.S. acquired the experimental data. C.P. and B.C. provided the nanoparticle samples. P.D. supervised the research. All team members contributed to the final revision of the article. 

\vspace{1cm}
{\noindent\Large{\textbf{Competing interests}}}\\\\
\noindent The authors declare no competing interests.

\end{document}

%% file: introduction.tex
\section{Introduction}
Improvements in electron microscopy (EM) techniques have been essential to advancing materials science \cite{Ophus2023, Lin2021} and molecular biology \cite{Gallagher-Jones2019, Lazic2022, Zhou2020} via imaging of objects at the nanoscale. Recent developments in direct electron detectors \cite{MacLaren2020-zg} have increased the capabilities of electron microscopy, enabling improvements in imaging \cite{Chen2021-dp, Nakane2020-jz}, diffraction \cite{Hattne2023,Saha2025}, and spectroscopy \cite{Hart2017-zq}. One type of detector, called an active pixel sensor (APS) \cite{Denes2007}, utilizes a complementary metal-oxide-semiconductor (CMOS) sensor with fast readout in order to reduce the number of electrons per frame to below about 1\% of the pixel number. Thus, each electron strike produces a localized deposition of energy that can be differentiated from other electron strikes and essentially ``counted'' as one electron regardless of the energy deposited~\cite{Battaglia2009-cn}. The process of converting an extended energy deposition to a single electron strike is known as electron counting. This process improves the signal-to-noise ratio relative to the raw sensor response and is central to EM experiments that involve direct electron detection based on the APS design \cite{Zhu2022,Peters2023}. The ``effective'' detector quantum efficiency can be increased with improved electron counting algorithms, producing high signal-to-noise frames at a wider range of fluences.

Counting electron strikes in APS detectors is challenging for several reasons. First, the amount of energy deposited in a thin volume of material by an energetic particle passing through it follows a Landau distribution \cite{Landau1944}. This captures both the most probable value of energy deposition, as well as a long tail extending to the full energy of the particle resulting from relatively rare interactions that leave behind large amounts of energy. For a large number of individual electron strikes in a given pixel the total energy deposition tends towards the most probable value, but for one to several hits, the Landau tail makes it difficult to quantify exactly the number of electrons that hit a pixel. Second, an electron can undergo multiple scattering in the medium, changing its direction of travel, producing clusters of energy responses with an inconsistent shape and extent \cite{PDG2024}. Finally, there is electronic noise \cite{McMullan2014} intrinsic to the CMOS hardware that can mask electron strikes with low energy deposition (false-negatives) or be incorrectly counted as electron strikes (false-positives). A counting algorithm that attempts to normalize the response to a single electron is therefore important in improving image quality. However, developing such an algorithm is challenging due to issues in segregating low-energy depositions from electronic noise and distinguishing multiple electron hits from single hits. These challenges are exacerbated by variations in the magnitude and shape of the energy deposition across the detector.

\begin{figure*}[!htb]
    \centering
\includegraphics[scale=0.54]{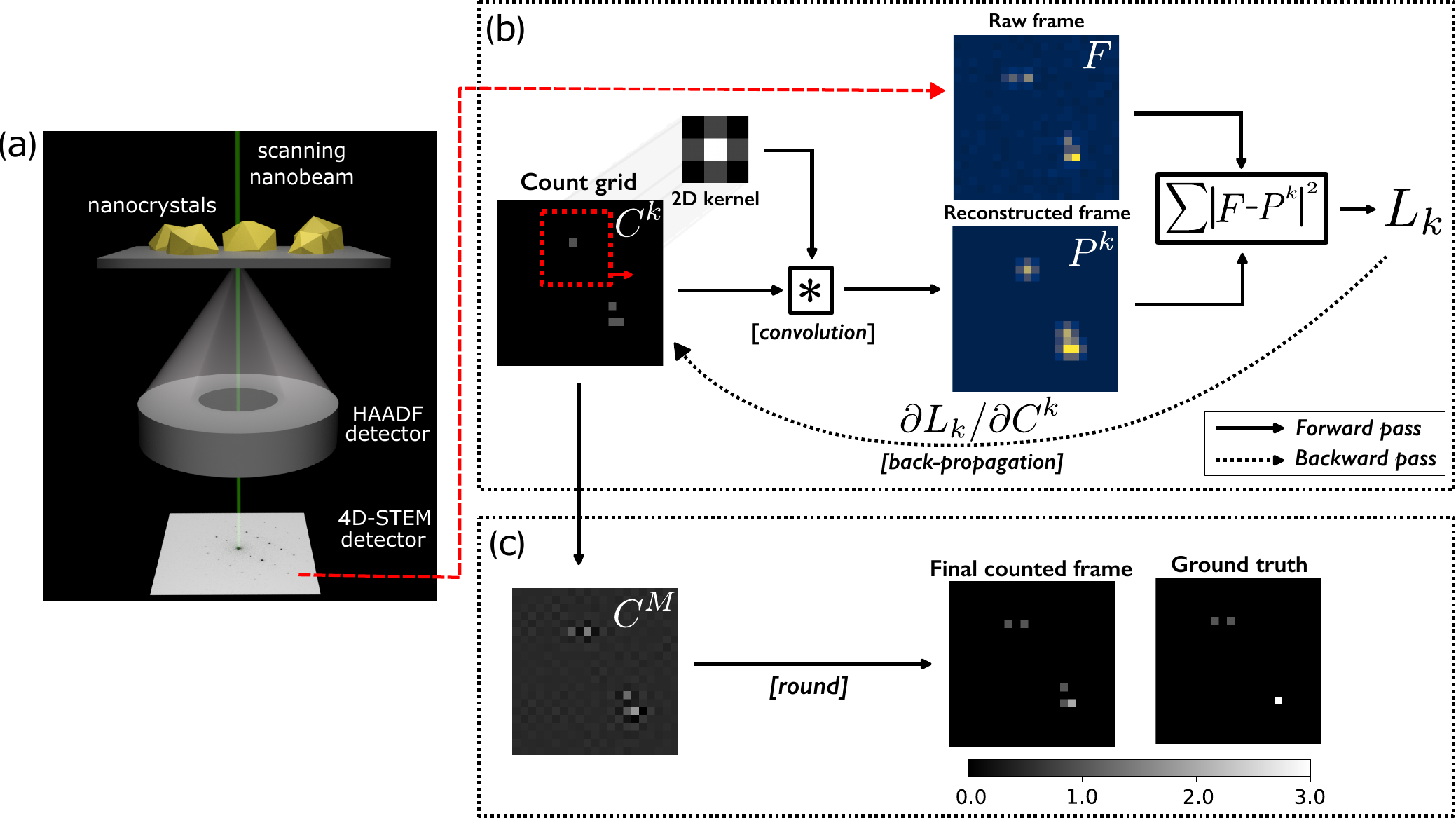}
\caption{\label{fig.method}\textbf{Illustration of 4D-STEM and Back-Propagation Counting (BPC).} (a) A scanning electron nanobeam incident on a sample produces a diffraction pattern captured by a pixelated APS-based detector. Simultaneously, high-angle scatterers are captured by a monolithic high-angle annular dark field (HAADF) detector. (b) An estimated counted frame is computed from the raw frame and convolved with the pre-computed 2D kernel to produce a reconstructed frame. Back-propagation is used to iteratively update the estimate according to a loss function. (c) The final optimized count grid is rounded to integer values. The ground-truth is shown for comparison but is not used in the algorithm.}
    \rule{\textwidth}{0.5pt}
\end{figure*}

APS sensors are being used to implement a new technique called four-dimensional scanning transmission electron microscopy (4D-STEM) \cite{Ophus2019}, a mode of operation capable of mapping crystallographic information with atomic- to nanoscale detail. In this technique, a focused electron probe is scanned over a sample in a two-dimensional (2D) raster pattern and a 2D diffraction pattern is measured at each probe position (see Figure \ref{fig.method}a). This results in a four-dimensional (4D) dataset with the potential to be hundreds of gigabytes to terabytes in size. 4D-STEM has become practical with the availability of fast, sensitive detectors leading to the need for computer hardware and algorithms to process such large datasets. To deal with this, electron counting algorithms are typically optimized for and built into specific camera hardware enabling in-line data processing during the experiment. Thus, a fast implementation of a counting algorithm is used (see \cite{stempy} for an open source implementation) which relies on simple thresholds to eliminate noise and energetic x-rays. These simple algorithms can only accommodate a low flux per frame on the camera and do not consider the shape of energy deposition over multiple pixels. 

Other approaches to electron counting use artificial intelligence, which has been increasingly employed in electron microscopy in recent years \cite{Botifoll2022, Treder2022, Gleason2024, Lobato2024}. In particular, convolutional neural networks (CNN) have been studied to improve localization accuracy and reduce coincidence loss \cite{vanSchayck2020, Wei2023}. However, the large numbers of parameters in these models, which must be optimized through back-propagation and stochastic gradient descent, require large amounts of simulated data and computation for training, and retraining is necessary for different detectors and electron energies. An approach that takes into account the energy deposition shape of individual electrons on the detector but does not require extensive detector/energy specific training could lead to substantial improvements that are broadly applicable, but are nascent \cite{Battaglia2009-cn}. 

To enable more accurate nanoscale imaging, we developed a novel method called Back-Propagation Counting (BPC) for electron counting. The algorithm makes use of modern machine learning techniques and software \cite{Pytorch} and back-propagation \cite{Rumelhart1986,Goodfellow2016} to fit the energy deposition of a raw 2D frame to a ``counted’’ frame of the same dimensions based on the expected average energy deposition of single electron strikes. The process does not involve training deep neural networks and thus does not require large amounts of training data. It can be readily applied with minimal understanding of the detector used in the acquisition. BPC processes low-occupancy frames with similar accuracy to standard methods and accommodates higher electron flux experiments with the ability to differentiate multiple electron hits per detector pixel. Through detailed characterization using synthetic and experimental data, we demonstrate that BPC results in quantitative and qualitative improvements in final data quality when compared to a standard counting method and removes artifacts. These results suggest that BPC offers improvements in materials imaging, especially in cases of high electron flux, enabling clearer reconstruction of material structures.

%% file: methods.tex
\section{Methods}\label{s.methodology}
The BPC method consists of the convolution of a Gaussian kernel over an estimated count grid, followed by optimization of the count grid values via back-propagation according to a squared error loss. It is essentially a ``deconvolution'' method implemented using back-propagation assuming a fixed response defined by the average single-electron energy deposition analogous to a ``point-spread function'' (PSF). A dataset of sparse electron hits on a given detector is helpful to characterize the average single-electron response, but the BPC method does not require a specific training phase or training data. Thus, the only important parameters that need to be specified can be estimated from either known detector characteristics (i.e thickness, pixel size, etc.) or a relatively small amount of sparse electron data. 

Figure \ref{fig.method} summarizes the BPC method workflow. Each electron is assumed to contribute its deposited charge in the shape of a 2D Gaussian with a specified amplitude and width. The parameters of the Gaussian are dependent on the detector and primary electron energy and remain fixed. Initially, the raw frame ($F$) is converted to an estimated initial count grid ($P^0$, see Figure \ref{fig.method}b), and the grid is optimized with stochastic gradient descent via back-propagation starting from this initial condition (described in more detail later). Each pixel in the count grid contains a floating-point value which is adjusted with each iteration in an attempt to match the raw frame. Once the optimization is complete, the pixel counts are rounded to the nearest integer. Additionally, we can account for the Landau fluctuations of the electron energy deposition in cases where a single electron strike may be erroneously identified as several due to the Landau tail. The method was implemented in PyTorch \cite{Pytorch} with the assistance of the large language model ChatGPT. Each step of the algorithm is described in more detail below.

\subsection{Single-electron Gaussian parametrization}\label{ss.gaussian}
The BPC method first requires the selection of appropriate parameters for the 2D Gaussian kernel, which estimates the single-electron energy deposition signature at the incident energy of interest. An approximate, though not necessarily optimal, determination of the amplitude $A_{e}$ and width $\sigma_{e}$ of this kernel can be extracted from a low electron flux dataset in which the vast majority of electron strikes are single-occupancy. A nominal single-electron threshold can be set, and patches of size $W \times W$ ($W$ odd), centered on pixels exceeding this threshold, can be averaged and fit to a 2D Gaussian,

\begin{equation}\label{eqn.gauss}
    G(A_{e},\sigma_{e}) = A_{e}\exp\Bigl(-\frac{x^{2} + y^{2}}{2\sigma_{e}^2}\Bigr).
\end{equation}

In this study, we use $W=3$ though the implementation of the method makes it straightforward to increase to $W=5$ or greater. In the synthetic data analyzed in Section \ref{ss.montecarlo}, we chose $A_{e} = 40$ and $\sigma_{e} = 0.5$, as discussed further in Section \ref{s.discussion}. For the experimental data in Sections \ref{ss.aperture} and \ref{ss.reconstruction}, we use $A_{e} = 19.03$ and $\sigma_{e} = 0.51$ which were extracted from a 2D Gaussian fit to an average single-electron pattern in the sparsest dataset (with lowest current) analyzed in Section \ref{ss.aperture}.

\subsection{Electron counting fit with back-propagation}\label{ss.counting}
Next, BPC reconstructs each raw frame by applying the 2D Gaussian determined in Section \ref{ss.gaussian} to a ``count grid'' where each pixel estimates a number of electron strikes. An initial count grid $C^0_{ij}$ is first estimated from the raw frame $F$ by subtracting a baseline $b$ based on the average dark noise and normalized by $A_e$. 

\begin{equation}
    C^{0}_{ij} = \Biggl\lfloor\frac{F_{ij} - b}{A_{e}}\Biggr\rceil.
\end{equation}

\noindent Here, $\lfloor \cdot \rceil$ refers to rounding to the nearest integer. The 2D Gaussian kernel is then convolved with the count grid to obtain a reconstructed frame,

\begin{align}\label{eqn.iteration}
    P^{k}_{ij} &= C^{k}_{ij} \ast G(A_{e},\sigma_{e}) \nonumber \\
     &= \sum_{l,m} C^{k}_{(i+l)(j+m)} G_{lm}(A_{e}, \sigma_{e}),
\end{align}

\noindent where $l$ and $m$ range from $-(W-1)/2$ to $(W-1)/2$. The elements of the count grid $C^{k}_{ij}$ are trainable parameters, and we denote the predicted frame as $P^{k}_{ij}$ and the measured frame as $F_{ij}$. We then compute the loss for the iteration, which we choose to be the sum of the squared differences of the pixels in the true and reconstructed frames,

\begin{equation}
    L_{k} = \sum_{i,j} (F_{ij} - P^{k}_{ij})^2.
\end{equation}

We then perform back-propagation with learning rate $\lambda$ (in this work we use PyTorch's implementation of the Adam optimizer with base $\lambda = 0.01$) to converge on a count grid that best reproduces the raw frame (Figure \ref{fig.method}b),

\begin{equation}
    C^{k+1} = C^{k} - \lambda\frac{\partial L_{k}}{\partial C^{k}}.
\end{equation}

By back-propagating the gradients through the loss $L_{k}$, we update the count grid for the next iteration to obtain $C^{k+1}_{ij}$. This strategy of updating the ``input'' via back-propagation rather than the parameters of the convolutional kernel is similar to that employed in \cite{Gatys2015} to produce images matching specific styles. Input optimization for a fixed model can also be found in studies of adversarial examples \cite{Goodfellow2015} in neural networks.

The procedure can be stopped after a given number of iterations $M$. Increasing $M$ results in a counted grid that more precisely matches the experimental data but obtaining a perfect match is not required for good counting. For the nanoparticle counting experiment discussed in Section \ref{ss.reconstruction}, we set a maximum of 3000 iterations. A dynamic stopping condition can also be implemented to activate when the loss decreases by less than a specified percentage over a set number of consecutive iterations. The final count grid is rounded to the nearest integer to obtain an integer number of electron counts for each pixel. 

\subsection{Addressing the Landau tail with a prior}\label{ss.prior}
As described earlier, the energy deposited by an electron in a thin detector will follow a Landau distribution. The long tail of this distribution indicates that a single electron can deposit, with relatively high frequency, significantly more energy than the most probable deposited energy. This will likely lead to many pixels being assigned high electron counting numbers, as BPC assumes an energy deposition shape with a fixed amplitude that will most likely be chosen to be within a factor of $\sim$2 of the most probable value in the Landau curve.

It is not possible to directly determine whether a given pattern of energy deposition belongs to one electron depositing much more energy than the most probable value or multiple electrons each depositing much less energy; however, we can use a statistical method to modify the optimized count grid in a way that aims to improve the counting. Here we assume that the electron occupancy across frames remains constant throughout the scan. The procedure is as follows:

\begin{enumerate}
    \item[1.] Construct a ``prior'' $\mu_{ij}$ which estimates the average number of electrons per frame striking each pixel in the counted dataset.
    \item[2.] Use the prior to compute the probability that, if a pixel is hit, it was hit by more than one electron.
    \item[3.] For all pixels in the optimized count grid $C_{ij}$ containing more than 1 hit count, reduce the hit count to 1 if a uniformly distributed random number is greater than the probability computed in step \#2.
\end{enumerate}

Note that the procedure makes no assumptions about the actual Landau distribution of energy deposition but rather removes cases of multiple hits when, based on the expected number of hits and Poisson statistics, they would be considered rare.

The prior $\mu_{ij}$ can be estimated from a number $N_{p}$ of frames $F_m$ as the average of these frames divided by a chosen characteristic single-electron amplitude. The value $A_{e}$ from equation \ref{eqn.gauss} can be used as this amplitude, and we require a minimum mean value of zero. Therefore, we have,

\begin{equation}\label{eqn.prior}
    \mu_{ij} = \max\left\{0, \Biggl(\frac{1}{N_{p}}\sum_{m=1}^{N_{p}} (F_{m})_{i,j}\Biggr) / A_{e}\right\}.
\end{equation}

Once the prior has been constructed, we take it to represent the mean number of electrons per frame striking each pixel. Assuming that this number follows a Poisson distribution, we can use it to compute the probability that $n_{ij}$ electrons strike pixel $(i,j)$ as $\mathrm{Poisson}(n_{ij};\mu_{ij})$. The quantity we are interested in for the purposes of reducing the counting error due to the Landau tail is the conditional probability that, given at least 1 electron hit, we have two or more hits, $P(n_{ij} \geq 2 | n_{ij} \geq 1; \mu_{ij})$. According to Bayes' theorem, 

\begin{align}
    & P(n_{ij} \geq 2 | n_{ij} \geq 1; \mu_{ij})\nonumber\\ 
    & =  \frac{P(n_{ij} \geq 1 | n_{ij} \geq 2; \mu_{ij})\cdot P(n_{ij} \geq 2; \mu_{ij})}{P(n_{ij} \geq 1; \mu_{ij})}\\
    & =  \frac{P(n_{ij} \geq 2; \mu_{ij})}{P(n_{ij} \geq 1; \mu_{ij})}\nonumber
\end{align}

\noindent since $P(n_{ij} \geq 1 | n_{ij} \geq 2; \mu_{ij}) = 1$. We can compute this conditional probability, knowing that the probability of drawing a Poisson number greater than some integer $l$ is equal to 1 minus the sum of all Poisson probabilities for numbers less than $l$, that is,

\begin{equation}
    P(n_{ij} \geq l; \mu_{ij}) = 1 - \sum_{m=0}^{l-1}\mathrm{Poisson}(m;\mu_{ij}).
\end{equation}

With this, we compute $P(n_{ij} \geq 2 | n_{ij} \geq 1; \mu_{ij})$ and compare it to a random number generated from a uniform distribution in the [0,1) interval for each pixel. For those pixels with more than 1 electron hit in the final counted grid $C_{ij}$, if the generated random number is greater than the computed probability, we set the value for that pixel to 1. In this way, improbable multiple hits are reduced to a single hit which was assumed to have deposited more energy due to the Landau tail. But for pixels with a high average number of electron hits, multiple hits are more likely to remain in the final counted grid.

This procedure does not address all possibilities. For example, it makes no attempt to determine whether a pixel with 3 apparent electron hits is actually 2 hits with 1 hit depositing above-average energy. A more sophisticated procedure would need to take into account the possible combinatorics of the composition of an $n$-electron hit response and the details of the Landau distribution, a task which is NP-hard and infeasible to confront in a practical application.

\subsection{Standard electron counting with stempy}\label{ss.stempy}
The standard electron counting algorithm used in this study is implemented in the stempy \cite{stempy} 4D-STEM analysis package. It takes as inputs the raw detector frame, a background threshold, and an X-ray threshold to exclude high-intensity events that were likely to have been the result of X-rays striking the detector. A dark reference frame for background subtraction and a gain map for pixel-wise sensitivity correction can also be provided as optional inputs.

The algorithm first applies the dark reference subtraction and gain correction (if provided) to each frame. The pixel intensities are then thresholded, so that pixels with values below the specified background threshold or above the X-ray threshold are excluded from consideration as electron hits. An electron strike is registered at a pixel's location if its intensity is greater than that of all of its eight immediate neighbors. In our analysis of simulated electron strikes (Section \ref{ss.montecarlo}), a background threshold of 10 counts and an X-ray threshold of 175 counts were employed.

\subsection{Simulation of electron hits in APS detectors} \label{ss.simulation}
To construct a 4D-STEM-like frame containing some number of electron hits, simulated hit patterns were placed on a blank grid such that their central pixel coincided with the pixel at the specified hit location. A Gaussian noise with a standard deviation of 1 count was added to all grid pixels. The hit patterns consisted of single-electrons with energy 100 keV incident on a silicon plate of 5 $\mu$m thickness and were simulated using Geant4 \cite{Geant4}. 

To model the charge collection process, diffusion was applied from the energy deposition sites to the edge of the plate. This was done by simulating a random walk for each electron using an estimated mean electron scattering length of 1 $\mu$m. The detection plane was binned to a 10x10 $\mu m^2$ pixel size, and the simulated pixel counts were then multiplied by a scaling factor so that the peak of the Landau distribution of the sum of all ADC counts in a single hit was approximately 20. 3 million single-electron hits were saved in a database for use in building multi-hit frames. In the simulation studies presented here, we consider frames for which electronic noise is insignificant, and therefore we add a low Gaussian noise of mean 0 and width of 1 to all pixels in frames constructed using these simulated hit patterns.

\subsection{Sample information} \label{ss.synthesis}
4D-STEM datasets were acquired of hexagonal-phase NaYF$_4$ core/shell nanoparticles with 15\% Tm$^{3+}$ core and 20\% Gd$^{3+}$ shell doping \cite{Pedroso2021,Qi2024}. These Tm$^{3+}$-based nanoparticles can both upconvert short-wave infrared light to shorter near-infrared wavelengths \cite{Qi2024} for use in imaging biological systems and be used as remote force sensors with exceptionally large dynamic range \cite{Fardian2025}. Their single-crystalline nature produces very strong Bragg peaks in diffraction patterns with a high probability of multiple electron strikes in detector frames.

\subsection{Electron microscopy}
Near parallel-beam 4D-STEM datasets were acquired using the TEAM 0.5 aberration corrected STEM \cite{Dahmen_2008} of the National Center for Electron Microscopy facility of the Molecular Foundry at Lawrence Berkeley National Laboratory. The STEM was operated at an accelerating voltage of 300 kV for all experiments. An APS detector called the 4D Camera \cite{Ercius2024} was used for the data collection with a frame rate of 87,000 Hz and 576 by 576 pixels per frame. The data shown in Figure \ref{fig.aperture_results}a were generated using a constant beam current (flux) of approximately 30 pA, and the post-specimen camera length was reduced to increase the current density on the camera. The process was also repeated with higher flux values to increase the likelihood of coincidence on the camera. The scan of nanoparticles presented in Figure \ref{fig.dp_and_rs} was acquired with a probe convergence angle of approximately 0.1 mrad and about 15 pA of beam current. 512 by 512 real space probe positions were used to image the 614 nm field of view with a probe step size of 1.2 nm. The size of the resulting raw dataset was 174 GB which was processed using the standard counting method and our newly developed BPC method. The BPC method processed frames in blocks of size 2064, each requiring of order 5 minutes to fit (approx. 7 frames/second) for 3000 iterations on an NVIDIA A100 GPU.

\subsection{Diffraction peak fitting}\label{ss.diffpeakfinding}
In Section \ref{ss.reconstruction}, we consider the intensities of the diffraction peaks corresponding to individual nanoparticles reconstructed in a 4D-STEM scan. The center of each nanoparticle was identified in the HAADF-STEM image for the scan, and a 2D circular mask of radius 6 pixels was used to extract the real-space coordinates corresponding to each one. By summing over the frames taken at the masked real-space coordinates, diffraction patterns for individual nanoparticles were constructed.

The various diffraction peaks in each pattern were fit to 2D Gaussians. Peaks were found in each pattern after eliminating all pixels below a chosen threshold value of 12. Only the most intense peak was kept when multiple peaks were identified within a minimum distance of 10 pixels, and only peaks common to both the BPC and standard reconstructions were considered.

%% file: results.tex
\section{Results}\label{s.results}
Here we present electron counting results using the new BPC method and compare them to a standard counting algorithm from \texttt{stempy} \cite{stempy}. We make use of both Monte Carlo simulation (Section \ref{ss.montecarlo}) and experimental data from two different studies (Sections \ref{ss.aperture} and \ref{ss.reconstruction}). 

\subsection{The BPC method counts electrons more consistently with increasing occupancy in synthetic data}\label{ss.montecarlo}
A goal of BPC was to succeed at counting over a wide range of electron occupancies where the standard method fails. To validate that BPC achieved this goal, we constructed simulated frames using Monte-Carlo-generated electron hits. This allowed us to study BPC under controlled conditions of pixel response amplitude and occupancy. We consider a case in which the charge deposited by a single electron is far greater than the noise present in the frames so that the results are independent of effects due to noise, and we focus specifically on counting in regimes of different electron occupancy (see Methods). 

To study electron counting under conditions of varying occupancy, electron hit locations were placed according to a Gaussian distribution with a width of $\sigma = 3$\footnote{Note that this $\sigma$ is the width of the Gaussian pattern and is unrelated to $\sigma_{e}$ from Eqn. \ref{eqn.gauss}.}, and the pixel responses for each hit were chosen randomly from the constructed database of simulated hit patterns (see Methods). A frame size of 51x51 pixels was chosen. A total of 10000 electron hits were distributed over $N_{MC}$ frames with a number of electrons per frame selected from a Gaussian distribution with mean $\chi$, such that $N_{MC}\cdot \chi = 10000$, in $(N_{MC},\chi)$ pairs of (10000, 1), (1000, 10), (100, 100), and (10, 1000). The counting was performed per frame, so the final electron statistics were approximately the same, but the counting took place under conditions of different occupancy. 

\begin{figure*}[!htb]
    \centering
    \includegraphics[scale=0.51]{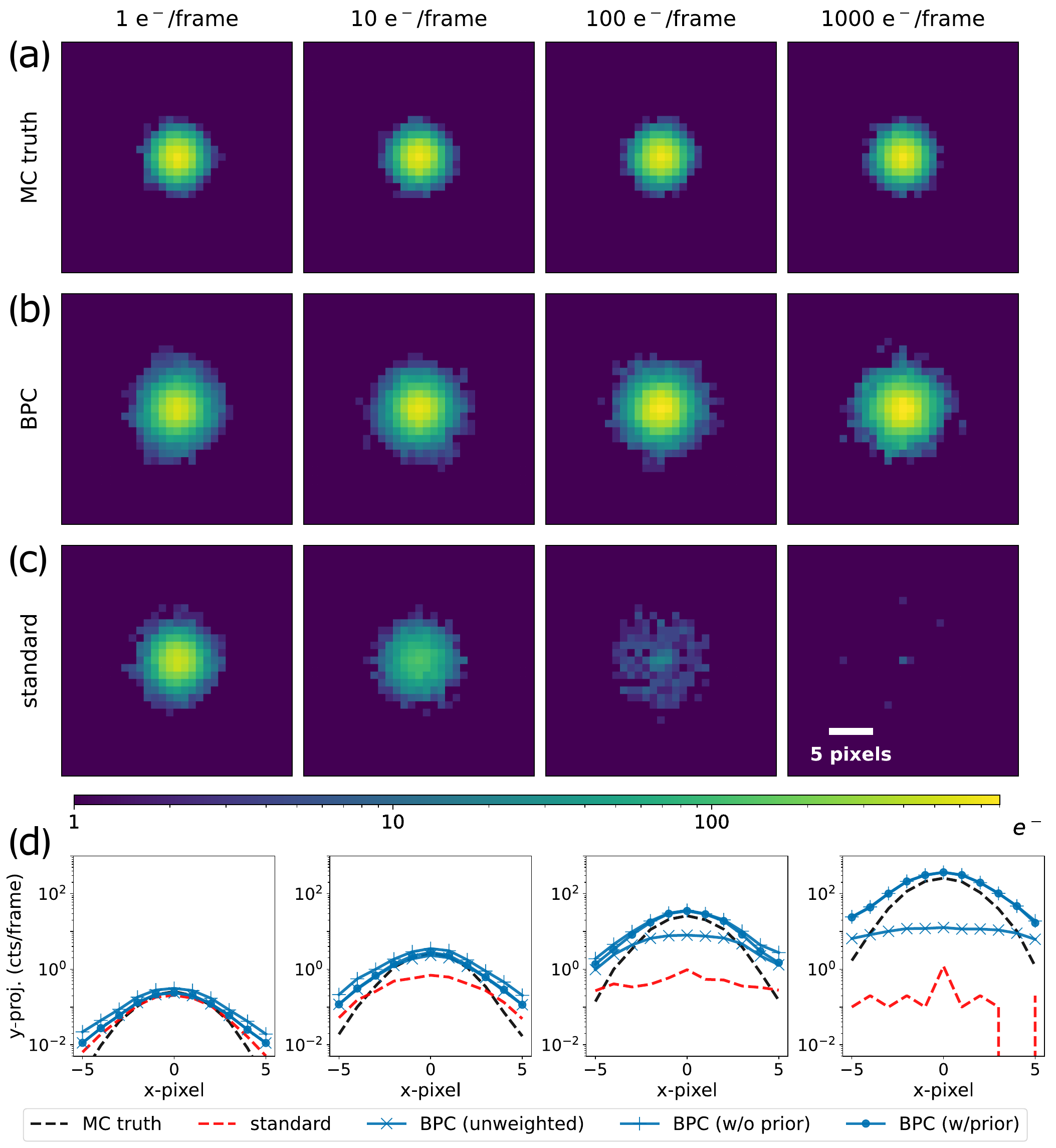}
    \caption{\label{fig.MC_counting}\textbf{The BPC method counts electrons more consistently with increasing occupancy in synthetic data.} A Gaussian ($\sigma = 3$ pixels) distribution of electron hits is simulated and counted for 1, 10, 100, and 1000 electrons per frame. The true number of electron hits summed over all frames in the Monte Carlo is shown in (a) for the different mean numbers of electron frames. The sums over all counted frames are shown using the BPC method in (b) and the standard method in (c). The final row (d) shows the counts projected along the y-axis, summed over all frames, for: the Monte Carlo truth, the standard method, a version of the BPC method with a maximum of 1 electron per hit (unweighted), the BPC method without application of the Landau correction described in Section \ref{ss.prior} (w/o prior), and the full BPC method (w/prior).}
    \rule{\textwidth}{0.5pt}
\end{figure*}

Figure \ref{fig.MC_counting} shows the simulated Monte Carlo Gaussian patterns and counting results for the different approximate numbers of electrons per frame. Figure \ref{fig.MC_counting}a shows the sum of the true electron hits for all Monte-Carlo-generated frames. Note again that while different numbers of electrons were incident on the frames shown in the four columns, the number of generated frames was varied so that the true number of hits was statistically similar in each case. Figure \ref{fig.MC_counting}b,c show the counted hits summed over all frames for the BPC and standard counting methods respectively. While both counting methods show an increased Gaussian width relative to truth (and BPC is indeed a little wider), BPC continues to accurately count at higher pixel occupancy where the standard method breaks down. To more directly visualize this, Figure \ref{fig.MC_counting}d shows the projections of the various 2D plots on the y-axis, including the MC truth from row \ref{fig.MC_counting}a, the standard method from row \ref{fig.MC_counting}c, and several variations on the BPC method: assuming all weights equal to 1 (“unweighted”), with application of weights but no attempt to correct for Landau fluctuations with the prior (``w/o prior''), and the full BPC method (``w/prior'') from row \ref{fig.MC_counting}b. From these results, we conclude that the new BPC method outperforms the standard method at higher occupancy in synthetic data, even without employing weights to account for multiple electrons per hit. 

\subsection{The BPC method consistently counts experimental data under a constant electron flux, evaluated over varying aperture sizes}\label{ss.aperture}
We next studied the performance of the BPC method under conditions of varying electron occupancy using experimental data. Unlike in Monte Carlo, we do not have the information of each true electron hit location. However, by using a fixed electron current and varying the post-specimen detector camera length, we can change electron density on the camera to test coincidence loss. Given the results from Section \ref{ss.montecarlo}, we hypothesized that BPC would correctly count electrons at higher density.

Experimental data at different camera lengths was collected at 300 kV and a set of fixed beam currents. 16512 frames of 576x576 pixels each were acquired for four different microscope camera lengths: 47.3 mm, 59.9 mm, 74.3 mm, and 92 mm. This was repeated for each of four different electron currents: $I_0 \approx$ 30 pA, $2I_0$, $4I_0$, and $8I_0$, for 16 datasets in total. Electron counting was performed using the standard method \cite{stempy} and using the BPC method. The first 1000 frames were used in constructing the prior for the Landau fluctuation corrections.

\begin{figure*}[!htb]
    \centering
    \includegraphics[scale=0.5]{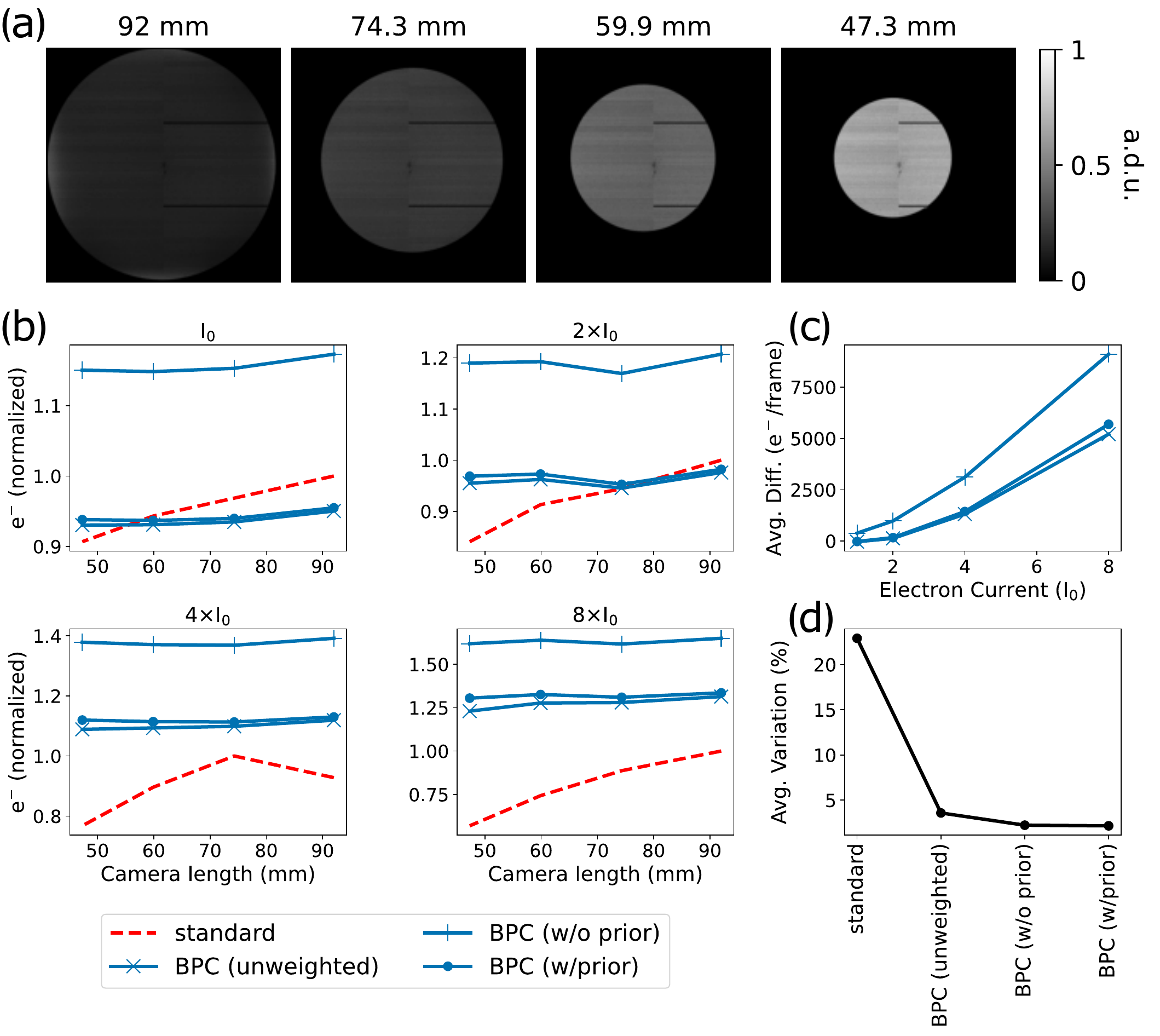}
    \caption{\label{fig.aperture_results}\textbf{The BPC method consistently counts experimental data under a constant electron flux, evaluated over varying aperture sizes.} The constant flux of electrons was measured over four different camera lengths (corresponds to aperture size) to provide a stable number of electrons with different occupancies. Similar datasets at electron currents $I_0 \approx 30$ pA, $2I_0$, $4I_0$, and $8I_0$ were acquired. (a) The circular aperture datasets acquired, summed over 16512 frames of 576x576 pixels, for electron current $8I_0$. (b) The total number of electron counts summed over 16512 frames vs. camera length for each of the 4 electron currents. The curves shown for each electron current were normalized to the maximum number of counts obtained using the standard counting method. (c) Average difference in the number of counts per frame between variations of the BPC method and the standard method, and (d) variation [(max-min)/max] in the curves shown in panel (b) for each method, averaged over all electron currents.}
    \rule{\textwidth}{0.5pt}
\end{figure*}

The results of the counting experiment are summarized in Figure \ref{fig.aperture_results}. The top panel of the Figure \ref{fig.aperture_results}a shows the raw data acquired for the highest electron current (approximately 240 pA) and the four different camera lengths. The increasing number of electrons per pixel with decreasing camera length is evidenced by the increasing intensity (color scale) from left-to-right. In Figure \ref{fig.aperture_results}b, for each electron current, the normalized sum of all counted electrons is shown for BPC (blue lines) and the standard method (red dashed line) for the four different camera lengths (counts were normalized to the maximum number of counts over all camera lengths obtained with the standard method for each current). As in Section \ref{ss.montecarlo}, three variations of the results of the BPC method are shown to better understand the effects of various components. All curves corresponding to the BPC method showed substantially reduced variations in the total number of counted electrons vs. camera length when compared to the standard method. For the lower currents, the number of electrons reported between the BPC and standard methods is similar, and at higher currents, BPC consistently reports higher count values. While the standard method does not account for an extended electron energy distribution or electron occupancy and assigns a weight of 1 to all counted pixels, BPC counts more consistently, as shown by the flatter normalized electron count vs. camera length curves even in the case in which one electron per counted pixel was assumed. This reflects an improved ability to count electrons that strike different nearby pixels, producing overlapping energy distributions. 

These observations are further quantified in Figure \ref{fig.aperture_results}c, which shows the difference in the counts/frame between BPC and standard methods for each electron current studied, averaged over the different camera lengths. We found that BPC counts more electrons than the standard method on average as the current, and therefore electron occupancy, increases. Finally, Figure \ref{fig.aperture_results}d shows the percentage variation [(max - min) / max] for each counting method in the plots shown in \ref{fig.aperture_results}b, averaged over the 4 different electron currents. As hypothesized, the percentage variation in the number of electrons counted at each camera length is reduced when employing the BPC algorithm, and in particular with the prior. This demonstrates the ability of BPC to more correctly count electrons crowded into a smaller region on the detector.

\subsection{The BPC method enhances diffraction from nanoparticles, yielding higher intensity diffraction peaks and improved image clarity}\label{ss.reconstruction}
Finally, we evaluated the performance of the method in identifying diffraction peaks corresponding to individual nanoparticles. Accurate reconstruction of single diffraction peaks, and in particular their intensities, is of critical importance in applications such as crystallography \cite{Borek2003}. Strong diffraction peaks are very likely to contain multiple electrons per frame, and thus the BPC method's ability to more accurately estimate the intensities of these peaks could lead to better crystallography metrics and structure solutions.

\begin{figure*}[!h]
     \centering
     \includegraphics[scale=0.45]{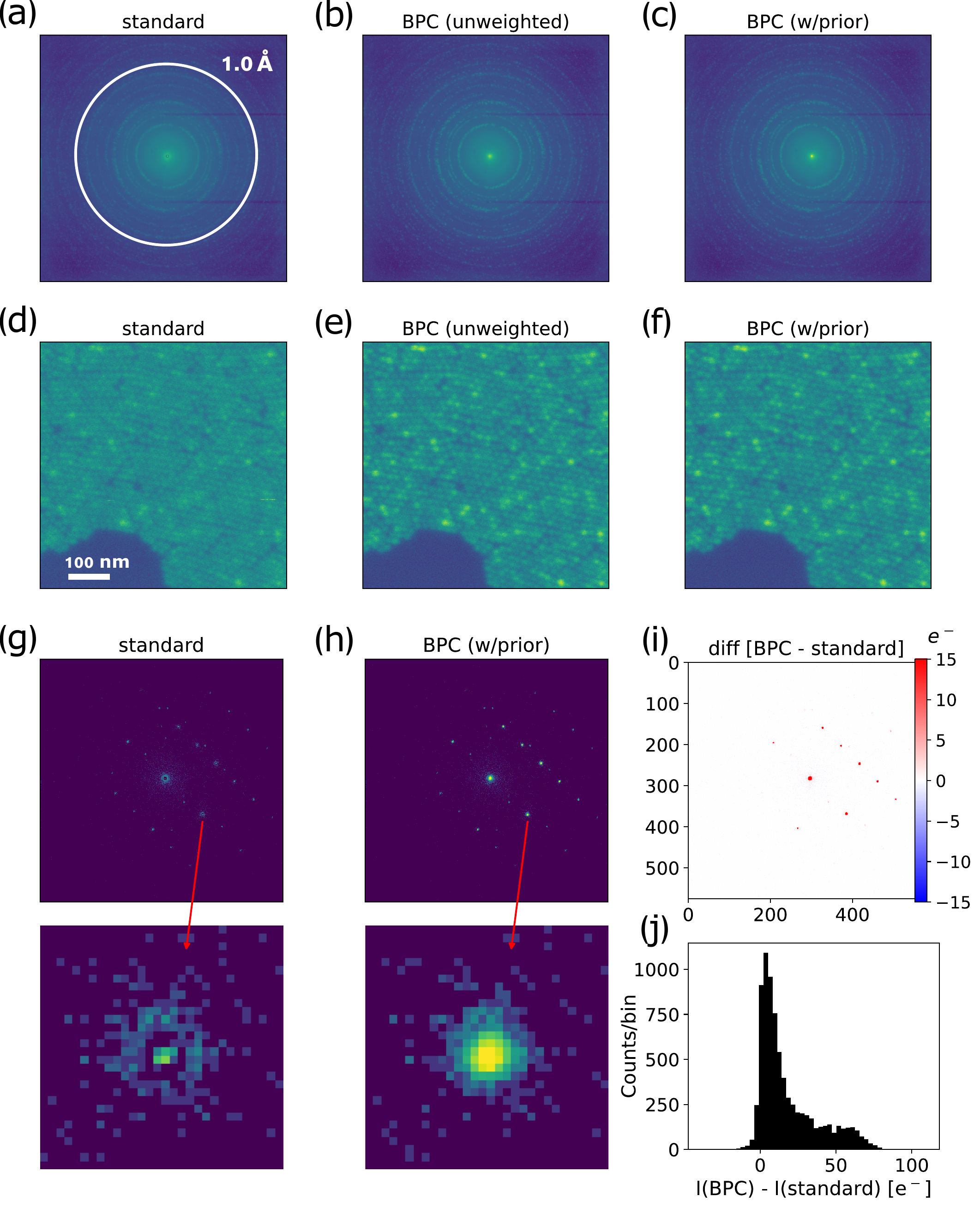}
     \caption{\label{fig.dp_and_rs}\textbf{The BPC method enhances diffraction from nanoparticles, yielding higher intensity diffraction peaks and improved image clarity.} The sum of counted diffraction patterns (a-c) and the real-space images constructed by summing the total diffraction pattern intensities at each probe position (d-f), as measured with a 4D-STEM detector and counted using the standard counting method and the BPC counting method, are shown. The diffraction patterns are shown with log-scale intensity, and all images in each row (a-c) and (d-f) are shown with equal contrast. An example diffraction pattern summed over real-space probe positions of a single nanoparticle, reconstructed with the standard (g) and BPC (h) counting methods, is also shown, as well as the difference $\mathrm{I}_{\mathrm{BPC}} - \mathrm{I}_{\mathrm{standard}}$ (i) between these two patterns. A single peak is highlighted to demonstrate a ``halo'' artifact observed in the standard counting for intense peaks. (j) Intensity difference for diffraction peaks selected from many nanoparticles, counted with the BPC and standard counting methods.}
\end{figure*}

The dataset consisted of 512x512 probe positions of doped NaYF$_4$ nanoparticles (see Methods). A monolithic HAADF detector was used to identify nanoparticles in real space. The raw frames, captured on the same detector as used in Section \ref{ss.aperture}, were counted with the standard method and the BPC method, employing 1000 frames to compute the occupancy prior for reduction of Landau over-counting. Figure \ref{fig.dp_and_rs} shows the summed diffraction patterns (Figure \ref{fig.dp_and_rs}a-c) and real space images showing the summed diffraction pattern intensity (excluding a region of radius 50 pixels about the zero-point) at each electron probe position (Figure \ref{fig.dp_and_rs}d-f), for which the counting was performed using the standard method (Figure \ref{fig.dp_and_rs}a,d) and the BPC method (Figure \ref{fig.dp_and_rs}b,e: unweighted, Figure \ref{fig.dp_and_rs}c,f: with occupancy weights and the application of the prior to correct for Landau fluctuations). We found that BPC yields visibly improved contrast and sharper diffraction peaks, even without employing occupancy weights, compared to the standard model.

The HAADF-STEM image acquired during the scan was used to identify individual nanoparticles in real space (see Methods).  A diffraction pattern corresponding to a single nanoparticle is shown in Figure \ref{fig.dp_and_rs}g for the standard counting method and in Figure \ref{fig.dp_and_rs}h for the BPC counting method. Various electron diffraction peaks are visible in the patterns. A single diffraction peak is identified with an arrow in Figure \ref{fig.dp_and_rs}g and \ref{fig.dp_and_rs}h and shown enlarged below the full diffraction patterns. A major effect that we found was that BPC eliminates a ``halo''-like reconstruction artifact present in the standard method in regions of high intensity. This effect is due to the final step of the standard method, which only permits assigning a count to pixels with intensities greater than their nearest neighbors, thus disallowing adjacent counts and possibly creating halo-like voids in regions of high intensity.

We directly compared the intensities of single diffraction peaks by taking the difference $\mathrm{I}_{\mathrm{BPC}} - \mathrm{I}_{\mathrm{standard}}$ of the two patterns (Figure \ref{fig.dp_and_rs}i). We found that the diffraction peaks were in many cases equally or more intense using BPC. To quantify this observation, we extracted peaks from several patterns (see Methods), and 7725 peaks were identified across 970 nanoparticle diffraction patterns. Figure \ref{fig.dp_and_rs}j shows the histogram of differences in intensities $\mathrm{I}_{\mathrm{BPC}} - \mathrm{I}_{\mathrm{standard}}$ for the common diffraction peaks that were found in both datasets. Critically, we found that the vast majority of peak intensities were larger for BPC compared to the standard method, sometimes by large amounts.

%% file: discussion.tex
\section{Discussion}\label{s.discussion}
Direct electron detectors are commonly deployed to image materials at the nanoscale. We addressed the problem of electron counting for direct electron detectors by developing the BPC method. We showed that BPC counts electrons more consistently with increasing occupancy in synthetic data. Analogously, in experimental data, we demonstrated the ability of BPC to more correctly count the same number of electrons even when they are crowded into a smaller region on the detector. Finally, in Nanobeam 4D-STEM diffraction experiments, we showed that the BPC method can produce diffraction patterns with visibly improved contrast and higher intensity diffraction peaks than the standard method. This is significant because the standard method is often used in 4D-STEM scans with APS detectors and was designed to work at low electron occupancy. Improved electron counting leads to more precise reconstructions of material properties—such as strain, composition, and local electromagnetic fields—critical for understanding nanoscale phenomena. Thus, our work paves the way for improved characterization and understanding of materials.

The BPC method could potentially be improved in future work. In particular, the 2D Gaussian single-electron response kernel could be generalized, for example by allowing its standard deviation (which is currently fixed) to vary. Furthermore, while the BPC method produces good results for the 4D Camera, which is back-thinned \cite{McMullan2009_backthinning}, it may be less suitable for electron hits measured in thicker CMOS detectors, in which the deposited charge pattern deviates significantly from a Gaussian distribution. Finally, we note that in the Monte Carlo frames described in Section \ref{ss.montecarlo}, over-counting was observed due to this effect when setting $A_{e}$ nearer to the value corresponding to the Landau peak value. In practice, $A_{e}$ can be adjusted to balance increased noise with detecting as many hits as possible.

In conclusion, we developed a novel electron counting algorithm for APS detectors based on convolutions with a pre-parameterized Gaussian kernel and back-propagation. The algorithm showed improved performance in counting electrons, in particular under conditions of increased pixel occupancy. It requires only the selection of two parameters ($A_{e}$ and $\sigma_{e}$), which can be estimated from a small detector calibration dataset. No data-intensive training step is necessary. While a faster standard counting method is most suitable for obtaining an initial first look at 4D-STEM data, the BPC method offers improvements, particularly in cases of higher electron occupancy as in the examples shown in this study, with significant implications for enhanced materials imaging.